\documentclass[letterpaper, 10 pt, conference]{ieeeconf}  

\IEEEoverridecommandlockouts                              

\usepackage{amsmath} 
\usepackage{amssymb}  
\usepackage{relsize}
\usepackage{bm}
\usepackage{float}
\usepackage{algorithm}
\usepackage{algpseudocode}
\usepackage{dutchcal}
\usepackage{cuted}
\usepackage{siunitx}
\usepackage[short]{optidef}
\usepackage{xcolor}
\usepackage{booktabs}
\usepackage{multirow}
\usepackage{colortbl}
\usepackage{array}
\usepackage{url}
\usepackage{svg}
\usepackage[11pt]{moresize}

\newcolumntype{H}{>{\setbox0=\hbox\bgroup}c<{\egroup}@{}} 

\usepackage[mode=buildnew]{standalone}

\title{\LARGE \bf
Adaptive Stochastic Nonlinear Model Predictive Control \\with Look-ahead Deep Reinforcement Learning \\for Autonomous Vehicle Motion Control}

\author{Baha Zarrouki$^{1,2}$, Chenyang Wang$^{2}$ and Johannes Betz$^{2}$
\thanks{$^{1}$ Chair of Automotive Technology, Technical University Munich}
\thanks{$^{2}$ Professorship of Autonomous Vehicle Systems, TUM School of Engineering and Design, Technical University Munich, 85748 Garching, Germany; Munich Institute of Robotics and Machine Intelligence (MIRMI), \{{baha.zarrouki}, {16chenyang.wang}, {johannes.betz}\}@tum.de
}}%

\begin{document}
\maketitle
\thispagestyle{empty}
\pagestyle{empty}
\begin{abstract}
In this paper, we present a Deep Reinforcement Learning (RL)-driven Adaptive Stochastic Nonlinear Model Predictive Control (SNMPC) to optimize uncertainty handling, constraints robustification, feasibility, and closed-loop performance. To this end, we conceive an RL agent to proactively anticipate upcoming control tasks and to dynamically determine the most suitable combination of key SNMPC parameters - foremost the robustification factor $\kappa$ and the Uncertainty Propagation Horizon (UPH) $T_u$. We analyze the trained RL agent's decision-making process and highlight its ability to learn context-dependent optimal parameters. One key finding is that adapting the constraints robustification factor with the learned policy reduces conservatism and improves closed-loop performance while adapting UPH renders previously infeasible SNMPC problems feasible when faced with severe disturbances. We showcase the enhanced robustness and feasibility of our Adaptive SNMPC (aSNMPC) through the real-time motion control task of an autonomous passenger vehicle to follow an optimal race line when confronted with significant time-variant disturbances. Experimental findings demonstrate that our look-ahead RL-driven aSNMPC outperforms its Static SNMPC (sSNMPC) counterpart in minimizing the lateral deviation both with accurate and inaccurate disturbance assumptions and even when driving in previously unexplored environments. 
\end{abstract}
\section{Introduction}
One approach to deal with poor closed-loop performance caused by uncertainties in Model Predictive Control (MPC) algorithms is to consider uncertainties in the MPC design and to guarantee the state and control constraints' satisfaction with a predefined probability, a concept known as Stochastic MPC \cite{rawlings2017model}.
Nevertheless, achieving real-time behavior with Stochastic Nonlinear Model Predictive Control (SNMPC) algorithms presents significant challenges. This stems from the difficulty in handling uncertainty propagation within nonlinear systems, frequently resulting in feasibility problems and a substantial computational overhead.\\
Various methods have been explored to address this issue, e.g. \cite{fagiano2012nonlinear} \cite{mesbah2014stochastic} \cite{bradford2021combining}. 
One recent approach \cite{zarrouki2023stochastic} introduces the concept of the Uncertainty Propagation Horizon (UPH) that limits the time for propagation of uncertainties through system dynamics. This promises to prevent infeasibility caused by uncertainty propagation divergence, optimizes closed-loop performance, and reduces computational times. This SNMPC approach uses Polynomial Chaos Expansion (PCE) techniques to propagate uncertainties effectively and consider nonlinear hard constraints on state expectations and nonlinear probabilistic constraints. 

\begin{figure}[h!]
  \centering
  \includegraphics[width=0.48\textwidth]{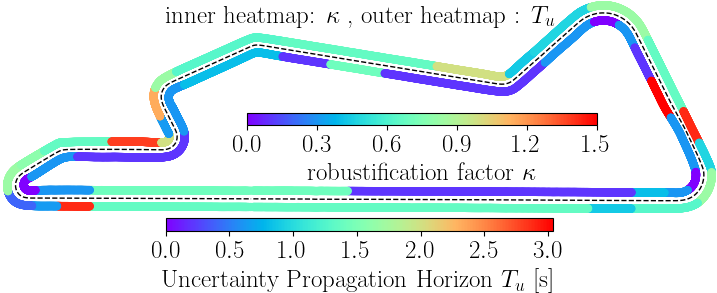}
  \caption{Reinforcement Learning agent's optimal decisions adapting two Stochastic NMPC parameters $\kappa$ and $T_u$ online for the motion control task for an autonomous vehicle on a racetrack. We see both parameters changing frequently to enhance optimal control behavior.}
  \label{fig:kappa_Tu_heatmap}
\end{figure}
However, this SNMPC approach remains sensitive to the length of the UPH and to the variance sensitivity factor $\kappa$ of the robustified nonlinear constraints to uncertainties. A bigger $\kappa$ means more consideration of the nonlinear constraints' variance caused by uncertainties and, thus, more tightened constraints. Choosing big $\kappa$ values when close to constraint limits, particularly when subject to significant disturbances, can yield an overly conservative control policy and degrade the closed-loop performance. In contrast, adopting small $\kappa$ values might result in frequent constraint violations. 
Furthermore, a large UPH may lead to a divergence in the simulated uncertain states' sequences and increased variances in nonlinear constraints, leading to infeasibility. In contrast, a small UPH may cause the controller not to account for state distributions adequately, thus compromising closed-loop performance.\\
In this paper, we propose to learn both parameters $\kappa$ and $T_u$ and automatically adapt them online with state-of-the-art Deep Reinforcement Learning (RL) techniques. Based on the current closed-loop performance assessment and future control reference trajectories, we design the RL agent to determine context-dependent optimal SNMPC parameters proactively. This promises a better closed-loop performance and an increased feasibility of dealing with a wide range of uncertainties and different dynamic situations.\\
Recent advancements have enabled RL agents \cite{sutton2018reinforcement} to effectively handle complex systems and to achieve performance levels comparable to human experts, not only in simulated environments but also in real-world applications \cite{nagabandi2018learning}\cite{song2023reaching}. RL algorithms push the agent to interact with its environment actively, facilitating the learning of a policy that optimizes a desired behavior by maximizing the rewards received as feedback.
Several works combine MPC with learning \cite{mesbah2022fusion}\cite{hewing2020learning}.
The authors of \cite{gros2020reinforcement} and \cite{cai2021mpcbased} employ MPC as a function approximator within RL. This approach ensures that the policy satisfies state and input constraints while meeting safety requirements.
Other learning-based MPC approaches use Bayesian Optimization \cite{mckinnon2018experience} \cite{wabersich2021probabilistic} or Gaussian Process Regression \cite{kabzan2019learning} \cite{bonzanini2021learning}. The authors of \cite{fiedler2022model} learn to adapt system dynamics from data collected during system operation. A sample-based learning MPC is proposed in \cite{rosolia2019sample} to approximate safe sets and the value function from historical data.
While RL is utilized as an end-to-end controller \cite{mnih2015human} \cite{levine2016end}, MPC is introduced as a safety filter for an RL-based controller \cite{wabersich2021predictive} \cite{wabersich2023data}. Furthermore, RL can also be used to learn controller design. Work \cite{bohn2021reinforcement} used RL algorithms to determine the prediction horizon of MPC to decrease computational complexity. Work \cite{bohn2023optimization} extends the previous work to learn other MPC meta-parameters that are non-differentiable wrt. the MPC outputs. Meanwhile, works \cite{zarrouki2021weights}\cite{zarrouki2020reinforcement} introduce a Deep RL driven Weights-varying NMPC to automatically learn and online adapt the cost function weighting matrices optimizing multi-control objectives.\\
One traditional adaptive SMPC \cite{oldewurtel2013adaptively}\cite{munoz2018stochastic} uses an adaptive law to iteratively update the chance constraints using a scaling factor that is proportional to the deviation of the empirical probability of violation from a desired violation level. 
The authors of \cite{santos2020stochastic} adjust individual chance constraints based on the empirical cumulative distribution function updated by online additive disturbance information, aiming for reduced conservatism. 
Prior research has predominantly concentrated on parameter adaptation in control systems to align with static reference profiles based on historical data. In contrast, our approach highlights the adaptation of SNMPC parameters to accommodate dynamic control references that change over time.
In motion control use-case, work \cite{suh2017new} adapts linear SMPC risk parameter using a cumulative distribution function and based on the current error between measured and predicted state without look-ahead capabilities. 
\\In summary, our work presents three main contributions:
\begin{enumerate}
\item We conceive an Adaptive SNMPC (aSNMPC) framework employing a look-ahead RL design. This agent autonomously adapts the SNMPC-specific parameters under varying uncertainties. 
\item We provide evidence of the increased robustness and feasibility of our aSNMPC compared to a standard SNMPC, particularly in the presence of fluctuating external disturbances. We conduct a comparative analysis of the effect of disturbance assumptions and illustrate the generalization capabilities of our aSNMPC when operating in previously unseen environments.
\item We assess the impact of learning and adapting each parameter on SNMPC performance, and we conduct a context-dependent analysis of the RL agent's decision-making process.
\end{enumerate}



\section{Stochastic NMPC}
 \label{sec:stochastic_nmpc}
In this work, we adopt the SNMPC framework proposed in \cite{zarrouki2023stochastic}. Problem 1 transforms a stochastic optimal control problem with probabilistic constraints into a deterministic one.
This framework aims to solve a stochastic OCP subject to chance constraints, offering an efficient method to transform chance constraints into robust deterministic constraints. Furthermore, it utilizes Polynomial Chaos Expansion (PCE) to propagate the uncertainties throughout the prediction horizon and proposes a novel UPH concept to address the infeasibility caused by uncertainty propagation. \\
In Problem 1, the following symbols are used: $\boldsymbol{x} \in \mathbb{R}^{n_x}$ represents the state vector, $\boldsymbol{u} \in \mathbb{R}^{n_u}$ represents the control vector, $T_p$ stands for the prediction horizon and $f$ denotes the system dynamics. The initial state is denoted as $\boldsymbol{x_0}$. Additionally, $l: \mathbb{R}^{n_{\mathrm{x}}} \times \mathbb{R}^{n_{\mathrm{u}}} \rightarrow \mathbb{R}$ defines the stage cost, while $m: \mathbb{R}^{n_{\mathrm{x}}}\rightarrow \mathbb{R}$ defines the terminal cost. \\
\begin{equation*}
\begin{aligned}
&\textbf{Problem 1} && \textbf{nominal Stochastic Nonlinear MPC}\\ 
\end{aligned}
\end{equation*}
\begin{equation}
\begin{aligned}
&  \underset{\substack{\boldsymbol{x}(.),\\ \boldsymbol{u}(.)}}{\min} & & 
\begin{aligned}
    \int^{T_p}_{\tau=0} & l( \mathbb{E}[\boldsymbol{x}(\tau)],\boldsymbol{u}(\tau)) \space  d\tau \\ 
 & + m(\mathbb{E}[\boldsymbol{x}(T_p)])
\end{aligned}
  \\ 
& \text{s. t.} & & \boldsymbol{x}_{0} \leq \boldsymbol{x}(0) \leq \boldsymbol{x}_{0} \text {, }  \\
& & & \dot{\boldsymbol{x}}(t) = f(\mathbb{E}[\boldsymbol{x}(t)],\boldsymbol{u}(t)) \text {, } & t \in[0, T_p), \\
& & & g(\mathbb{E}[\boldsymbol{x}(t)], \boldsymbol{u}(t)) \leq \bar{\boldsymbol{g}},& t \in[0, T_p), \\
& & & 
\begin{aligned}
     \mathbb{E}[h(\boldsymbol{x},\boldsymbol{u})] + \kappa\sqrt{\text{Var}[h(\boldsymbol{x},\boldsymbol{u})]}
\leq \bar{\boldsymbol{h}},
\end{aligned}
& t \in[0, T_p), \\
& & &  J_{\mathrm{bx}} \space \mathbb{E}[\boldsymbol{x}(t)] \leq \bar{\boldsymbol{x}}, & t \in[0, T_p), \\
& & & J_{\mathrm{bu}} \space \boldsymbol{u}(t)\leq \bar{\boldsymbol{u}}, & t \in[0, T_p), \\
& & & g^{\mathrm{e}}(\mathbb{E}[\boldsymbol{x}(T_p)]) \leq \bar{\boldsymbol{g}}^{\mathrm{e}}, \\
& & & 
\begin{aligned}
\mathbb{E}[h^{\mathrm{e}}&(\boldsymbol{x}(T_p))] \\&+ \kappa\sqrt{\text{Var}[h^{\mathrm{e}} (\boldsymbol{x}(T_p))]}
\leq \bar{\boldsymbol{h}}^{\mathrm{e}},
\end{aligned}
\\
& & & J_{\mathrm{bx}}^{\mathrm{e}} \space \mathbb{E}[\boldsymbol{x}(T_p)]\leq \bar{\boldsymbol{x}}^{\mathrm{e}} \\
\end{aligned}
\label{eq:SNMPC problem}
\end{equation}
The SNMPC in Problem 1 incorporates hard linear constraints on the states' expectations and on the control inputs formulated with help of $J_{bx}$, $J_{bx}^e$ and $J_{bu}$. Furthermore, it handles hard nonlinear constraints on the states' expectations: $g$ and $g^{\mathrm{e}}$. Problem 1 transforms nonlinear probabilistic inequality constraints into estimated deterministic surrogates in expectation and variance of the nominal path and terminal nonlinear inequality constraints: $h$ and $h^e$. Here, $\kappa = \sqrt{(1- p)/p}$ denotes the nonlinear constraints variance sensitivity factor, i.e. constraints robustification factor, meant to tighten the constraints according to the variance of the nonlinear constraints affected by uncertainties. Here, $p \in (0,1]$ is the desired probability of violating the nonlinear constraints $h$.
\\The expectation and variance of the states and nonlinear constraints are estimated with Polynomial Chaos Expansion (PCE) method through propagating $n_s$ sampled points around the current measured state to account for uncertainties as in Eq.\ref{eq:propagation of states and constraints}. The propagation of uncertain state samples and constraints through $T_p$ is limited by Uncertainty Propagation Horizon (UPH): $T_u$. After reaching the UPH, the propagation of the samples is stopped and only the last estimated variables at $t = T_u$ are propagated until $T_p$. 
\begin{equation}
    \begin{cases}
    \begin{aligned}
        &\mathbb{E}[\boldsymbol{x}_{t}]= \boldsymbol{c}^{(\boldsymbol{x})}_0\\
        &\mathbb{E}[h(\boldsymbol{x},\boldsymbol{u})] = c^{(h)}_0\\
        &\text{Var}[h(\boldsymbol{x},\boldsymbol{u})] = \sum_{k=1}^{L-1}
    (c^{(h)}_k)^2  
    \end{aligned}
    &\text{, if } t \in \{0,...,N_{u-1}\}  \\ \\
     \begin{aligned}
        &\mathbb{E}[\boldsymbol{x}_{t}]= \boldsymbol{x}_{t} = f(\mathbb{E}[\boldsymbol{x}_{t-1}],\boldsymbol{u})\\
        &\mathbb{E}[h(\boldsymbol{x},\boldsymbol{u})] = h(\mathbb{E}[\boldsymbol{x}_t],\boldsymbol{u})\\
        &\text{Var}[h(\boldsymbol{x},\boldsymbol{u})] = 0
    \end{aligned}
    &\text{, if } t \in \{N_{u},...,N_{p-1}\}  \\
    \end{cases}
    \label{eq:propagation of states and constraints}
\end{equation}
Here, $c^{(h)}_k$ and $\boldsymbol{c}^{(\boldsymbol{x})}_0$ represent the PCE coefficients of the nonlinear inequality constraints and the states respectively, $L$ the total number of the PCE terms and $N_{u}$ and $N_{p}$ denote the number of shooting nodes within the UPH and prediction horizon respectively. For further details, we refer to \cite{zarrouki2023stochastic}.

\section{Stochastic NMPC for Trajectory Following of Autonomous Vehicles}
\label{sec:traj_following}
We conceive the SNMPC to control the longitudinal and lateral motion of our Volkswagen T7 Multivan autonomous research vehicle \cite{karle2023edgar} to follow a given trajectory while being subjected to state estimation uncertainties: $\boldsymbol{x}_{t+1} = f(\boldsymbol{x}_{t},\boldsymbol{u}) + \boldsymbol{w}_t$. Here, $\boldsymbol{w}_t$ is the disturbance and $\boldsymbol{x} = [x_{\text{pos}},\space y_{\text{pos}},\space \psi,\space v_{\text{lon}},\space v_{\text{lat}},\space \dot{\psi},\space \delta_f,\space a]^T $ is the state vector with the yaw angle $\psi$, the yaw rate $\dot{\psi}$, the steering angle at the front wheel $\delta_f$ and the acceleration $a$. The control vector $\boldsymbol{u} = [j, \space \omega_f ]^T$, where $j$ is the longitudinal jerk, and $\omega_f$ is the steering rate. We adopt a dynamic nonlinear single-track model as our prediction model. We refer to \cite{zarrouki2023stochastic} for the full dynamics definition.
We define the stage- and terminal costs as $l(\boldsymbol{x}, \boldsymbol{u})=\frac{1}{2}\|\boldsymbol{y}(\boldsymbol{x},\boldsymbol{u})-\boldsymbol{y}_{\mathrm{ref}}\|_W^2$ and $m(\boldsymbol{x})=\frac{1}{2}\|\boldsymbol{y}^e(\boldsymbol{x})-\boldsymbol{y}^e_{\mathrm{ref}}\|_{W^e}^2$ with $W$ and $W_e$ being the weighting matrices.

The system is subject to combined longitudinal and lateral acceleration limits that we formulate using the transformed nonlinear chance constraints, i.e. robustified constraints:
\begin{equation}
\begin{aligned}
     &\mathbb{E}[h(\boldsymbol{x},\boldsymbol{u})] + \kappa\sqrt{\text{Var}[h(\boldsymbol{x},\boldsymbol{u})]}
\leq 1\\
    &h(\boldsymbol{x}, \boldsymbol{u}) = \left(\frac{a_{\text{lon}}}{a_{x_\text{max}}}\right)^2 + \left(\frac{a_\text{lat}}{a_{y_\text{max}}}\right)^2
    \end{aligned}
    \label{eq:combined constraints}
\end{equation}
Furthermore, we formulate linear hard constraints on the steering angle expectation $\delta_f$ and on the steering rate control input at the front wheel $\omega_f$:
\begin{equation}
\begin{aligned}
& |\mathbb{E}[\delta_f]| \leq 0.61\unit{\radian} \\
&|\omega_f| \leq 0.322\unit{\radian\per\second}
\end{aligned}
\label{eq:hard constraints}
\end{equation}
\label{SNMPCconfiguration}
We assume that the predicted states $v_\text{lon}$, $v_\text{lat}$ and $\dot{\psi}$ are subject to uncertainties. We assume that the uncertainties are Gaussian disturbances and the correct standard deviations are known, such as:
\begin{equation}
\begin{aligned}
    \boldsymbol{\sigma}_{w}^{\text{SNMPC}} = [\sigma_{\text{vlon}}, \sigma_{\text{vlat}}, \sigma_{\dot{\psi}}] ^T  = [\sigma_{\text{vlon}}^{\text{sim}}, \sigma_{\text{vlat}}^{\text{sim}}, \sigma_{\dot{\psi}}^{\text{sim}}]^T
\end{aligned}
\end{equation} 
Once integrated into the vehicle, the state estimation module computes the standard deviations relying on sensor data.

\section{Learning SNMPC parameters with Look-ahead Deep Reinforcement Learning}
\label{sec:rl_snmpc}
\begin{figure}[H]
  \centering
  \includegraphics[width=0.48\textwidth]{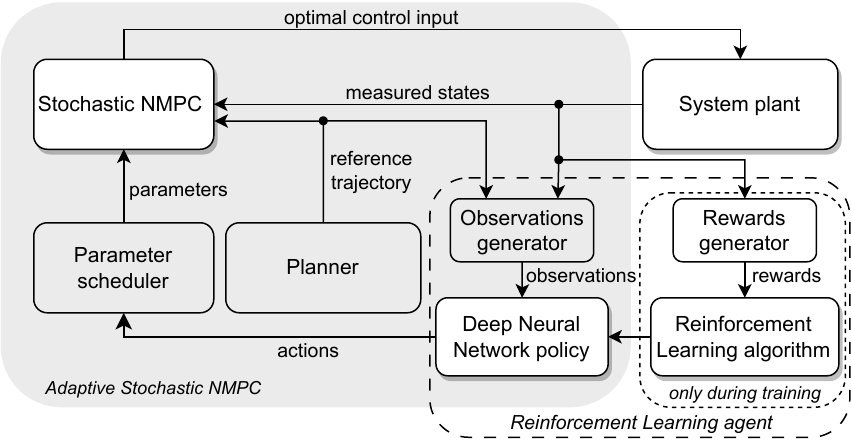}
  \caption{Architecture of the Deep RL driven aSNMPC}
  \label{fig:architecture of DRL-SNMPC}
\end{figure}

In this section, we present our Adaptive SNMPC (aSNMPC), which automatically adjusts the nonlinear constraints robustification factor $\kappa$ and the UPH $T_u$ according to changing environments and driving tasks. We leverage a Deep Neural Network (DNN) policy, learned through state-of-the-art Deep RL (DRL) algorithms, to dynamically tune these parameters. Fig.\ref{fig:architecture of DRL-SNMPC} illustrates the architecture of our DRL-driven aSNMPC. 
The DNN policy computes new actions (Sec.\ref{subsec:action space}), and the parameter scheduler alters the SNMPC parameters every predefined switching time instance $T_{sw} = n\cdot T_\text{s,sim}, n \in \mathbb{N}$, where $T_\text{s,sim}$ is the environment discretization time. The new actions are computed with a feed-forward step through the DNN using observations generated based on current measured states and the future reference trajectory (Sec.\ref{subsec:observation space}). \\
During the training/learning phase, the RL algorithm updates the DNN policy based on the collected rewards (\ref{subsec:designing reward function}) after a certain number of steps in the environment $n_\text{steps}$, i.e., new updated SNMPC parameters steps. After the training is done, the rewards generator, as well as the RL algorithm, are no longer needed. The rest of the architecture remains the same for the deployment phase. 
In this work, we adopt the Proximal Policy Optimization (PPO) method \cite{schulman2017proximal} as our RL algorithm. PPO has gained attention for its ability to optimize policies with minimal hyperparameter adjustments.
Figure \ref{fig:NN_architecture} delineates the architecture of the DNN policy we conceive for this problem.
\subsection{Defining the action space}
\label{subsec:action space}
The UPH is within the propagation horizon, i.e. $T_u \in [0,T_p]$. Given that $T_u \in \mathbb{N}$, we choose a discrete action space which comprises $N_p = \frac{T_p}{T_s}$ discrete steps, each corresponding to a shooting node of the SNMPC. \\
To maintain a standardized action scale for constraint robustification, we segment $\kappa$ into 21 discrete intervals, spanning from $\kappa_\text{min}=0$ to $\kappa_\text{max}=2$ at intervals of 0.1. Experiments have shown that RL agents choose only robustification factor values less than 2. 
\subsection{Designing the reward function}
\label{subsec:designing reward function}
As the cost function already optimizes the trajectory following of the vehicle, in this work, we design our DRL agent to optimize the SNMPC's performance by minimizing the following aspects:
\begin{enumerate}
    \item  infeasibility
    \item  constraint violations
    \item  lateral deviation 
\end{enumerate}
Accordingly, we give a high reward for small lateral deviations, no reward when the constraints are violated and we impose a penalty when the SNMPC can't find a solution under the chosen parameters. Hence, the reward function is defined as following:
\begin{equation}
R=
\begin{cases}
\begin{aligned}
    &-A ,
    &&\text{if SNMPC is infeasible}  \\
    &0 , 
    &&\text{if constraints are violated} \\
    &A \cdot exp\left( - \frac{e_\text{lat}}{\sigma_\text{lat}}\right),
    &&\text{else } \\
\end{aligned}
\end{cases}
\end{equation}
Here, $A$ represents the peak of reward, i.e., the maximum reward fed back to the agent, and $\sigma_\text{lat}$ represents the spread in $e_\text{lat}$. Notably, $e_\text{lat}$ represents the lateral deviation computed based on the undisturbed states in simulation. When $e_\text{lat} = 0$, the agent gets the highest reward $A$. 
In this work, we adopt $A=1$ and $\sigma_\text{lat} = 1\unit{\meter}$ in this work.
Note that 
$e_\text{lat} = \max [|e_{\text{lat},1}|, |e_{\text{lat},2}|, \dots, |e_{\text{lat},n}|]$ with $n = \frac{T_{sw}}{T_\text{s,sim}}$, i.e. the absolute maximum measured lateral deviation between two parameter switching intervals $T_{sw}$.
\subsection{Defining the observations}
\label{subsec:observation space}
\begin{figure}[t]
  \centering
  \includegraphics[width=0.47\textwidth]{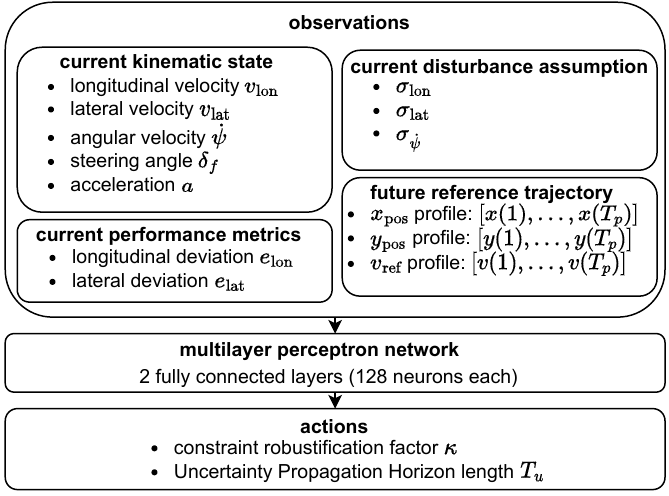}
  \caption{Model of the Deep Neural Network driven aSNMPC}
  \label{fig:NN_architecture}
\end{figure}

The observation space represents the agent's perceptual input from the environment. We design it to provide the agent with a comprehensive set of relevant information necessary for informed decision-making as in Fig.\ref{fig:NN_architecture}. We make the agent access the current ego kinematic state while decoupling it from specific coordinates, a design choice aimed at enhancing the generality of the learned policy. Additionally, the agent continually monitors its performance by assessing the outcomes of its most recent actions. \\
Given that the SNMPC relies on the assumption of uncertainty distributions, the agent must have access to these assumptions. When dealing with larger uncertainties, the likelihood of constraint violations increases, resulting in a smaller robustification factor $\kappa$. This, in turn, requires a reduced UPH to maintain system stability.\\
Lastly, we equip the agent with the capability to anticipate the future by providing it with the current reference trajectory within the prediction horizon. This foresight enables the agent to proactively determine the optimal parameters that align with the specified objectives, guided by the anticipated dynamic profiles that the SNMPC should follow. The agent receives detailed future profiles, including $x$ and $y$ positions as well as velocity. To allow for generality, we disassociate the $x$ and $y$ coordinates from the map by transforming the reference trajectory to the ego vehicle's coordinates. This allows the agent to make informed decisions regarding future combined longitudinal and lateral behaviors, distinguishing between scenarios like high-speed curves and straights or curves with substantial lateral uncertainty.
Based on the discussions above, we define a general algorithm that learns and adapts the SNMPC parameters with RL (Algorithm 1). 

\begin{algorithm}[ht]
\caption{RL driven Adaptive-SNMPC}\label{alg:RL_SNMPC}
\algrenewcommand\algorithmicrequire{\textbf{Init}}
\begin{algorithmic}[1]
\State Set SNMPC parameters: $T_p$, $T_{s}$, $T_u$, $\kappa$, $n_{s}$, $\boldsymbol{\sigma}_{w,\text{SNMPC}}$, $W$ and $W_e$
\For{$i \in \left\{1,...,N\right\}$}
    \State Update initial state $\boldsymbol{x}_0$ with current measurements
    \State Generate samples around the current initial state $\boldsymbol{x}_0$
    \State Compute the PCE matrix solution 
    \State Update current SNMPC reference
    \If{\text{switching time (}$i \bmod T_{sw} = 0$)}
		\State Generate observations and rewards
		\State Execute the current RL policy
		\State Update SNMPC parameter $\kappa$ and $T_u$
        \If{learning-step ($i \bmod n_\text{steps}\cdot T_{sw} = 0$)}
        \State Execute RL policy update
        \EndIf
    \EndIf
    \For{$j \leq N_p$}
        \State Propagate the uncertain states $\mathbb{E}[\boldsymbol{x}_{j}]$ (Eq.\ref{eq:propagation of states and constraints}) 
        \State Estimate nonlinear constraints $\mathbb{E}[h_j]$ and $\text{Var}[h_j]$
    \EndFor
    \State Solve the SNMPC problem (Eq.\ref{eq:SNMPC problem})
    \State Apply the first control input $\boldsymbol{u}^*_0$ on the real system
\EndFor
\end{algorithmic}
\end{algorithm}

\section{Simulation Results: RL driven Adaptive SNMPC}
\label{sec:sim_results}
We conduct a performance comparison between our Adaptive SNMPC (aSNMPC) approach and its static counterpart, Static SNMPC (sSNMPC) \cite{zarrouki2023stochastic}. Both are subjected to significant additive Gaussian disturbances that severely impact the quality of state estimates. The SNMPC configurations are detailed in Table \ref{tab:SNMPC parameter}, while our SNMPC implementation specifics can be found in \cite{zarrouki2023stochastic}.\\
The RL agent's training is carried out on an AMD Ryzen 7950X 5.70 GHz CPU using the Proximal Policy Optimization (PPO) algorithm implemented within the Stable Baselines3 framework \cite{stable-baselines3}. Configuration details are outlined in Table \ref{tab:PPO parameter}. The training process in our simulation environment spans approximately 22 hours.
\begin{table}[H]
\centering
\caption{Hyper parameters of the Proximal Policy Optimization}
\label{tab:PPO parameter}
\begin{tabular}{ll}
\toprule
\textbf{Parameter}&\textbf{Value} \\
\midrule
Learning rate $\alpha$&0.0007 \\
Policy update every $n_\text{steps}$&512 \\
Discount factor $\gamma$ & 0.99\\
GAE factor $\lambda$ & 0.95\\
Clipping parameter $\epsilon$ & 0.2\\
Training steps $N$ & $10^{6}$\\
\bottomrule
\end{tabular}
\end{table}
We design the RL agent with the following termination conditions: an episode terminates when the agent completes one lap.  Furthermore, the episode truncates if the SNMPC problem becomes infeasible. After each episode, the standard deviations of the simulated disturbances are randomized, and their values are known to the agent. We train the agent in a racetrack environment, and we evaluate its capability of generalization on other unseen racetracks (Sec.\ref{subsec: generalization}). All the experiments in the following sections are evaluated by letting the MPC control the vehicle for $110\unit{\second} = \frac{110}{T_\text{s,sim}} = 5500$ MPC steps.
\label{subsec:simulation_setup}
\begin{table}[h]
\caption{Standard deviation ranges used to sample and simulate changing disturbance distributions}
\label{tab:std ranges}
\centering
\begin{tabular}{lll}
\toprule
  \text{$\sigma$} & $\sigma_\text{min}$ & $\sigma_\text{max}$ \\
\midrule
 $\sigma_{x,y} [\unit{\meter}]$ & 0.1 & 0.3 \\ 
 $\sigma_{\psi} [\unit{\radian}]$ & 0.008 & 0.017 \\ 
 $\sigma_{v,\text{lon}} [\unit{\meter\per\second}]$ & 0.5 & 1.0 \\ 
 $\sigma_{v,\text{lat}} [\unit{\meter\per\second}]$ & 0.5 & 1.0 \\ 
 $\sigma_{\dot{\psi}} [\unit{\radian\per\second}]$ & 0.04 & 0.08 \\ 
$\sigma_{\delta_f} [\unit{\radian}]$ & 0.001 & 0.0017 \\
\bottomrule
\end{tabular}
\end{table}
\begin{table}[H]
\centering
\caption{SNMPC parameters and disturbance configuration}
\label{tab:SNMPC parameter}
\begin{tabular}{ll}
\toprule
\textbf{Parameter}&\textbf{Value}\\
\midrule
Simulation sampling time $T_\text{s,sim}$ & 0.02\unit{\second} \\
SNMPC discretization time $T_s$ & 0.08\unit{\second} \\
Prediction horizon $T_p$ & 3.04\unit{\second} \\
Default UPH $T_u$ & 2\unit{\second} \\
Default robustification factor $\kappa$ & 0.42\\
Parameter switching time $T_{sw}$ & $0.8 \cdot T_p$ \\
Disturbance ranges switch $T_{\boldsymbol{\sigma},sw}$ & 30 \unit{\second} \\
Simulation duration per evaluation& 110\unit{\second}  \\
\bottomrule
\end{tabular}
\end{table}
\subsection{Trajectory Following Performance}
\label{subsec:traj_foll_performance}
\begin{figure}[H]
\includegraphics[width=.48\textwidth]{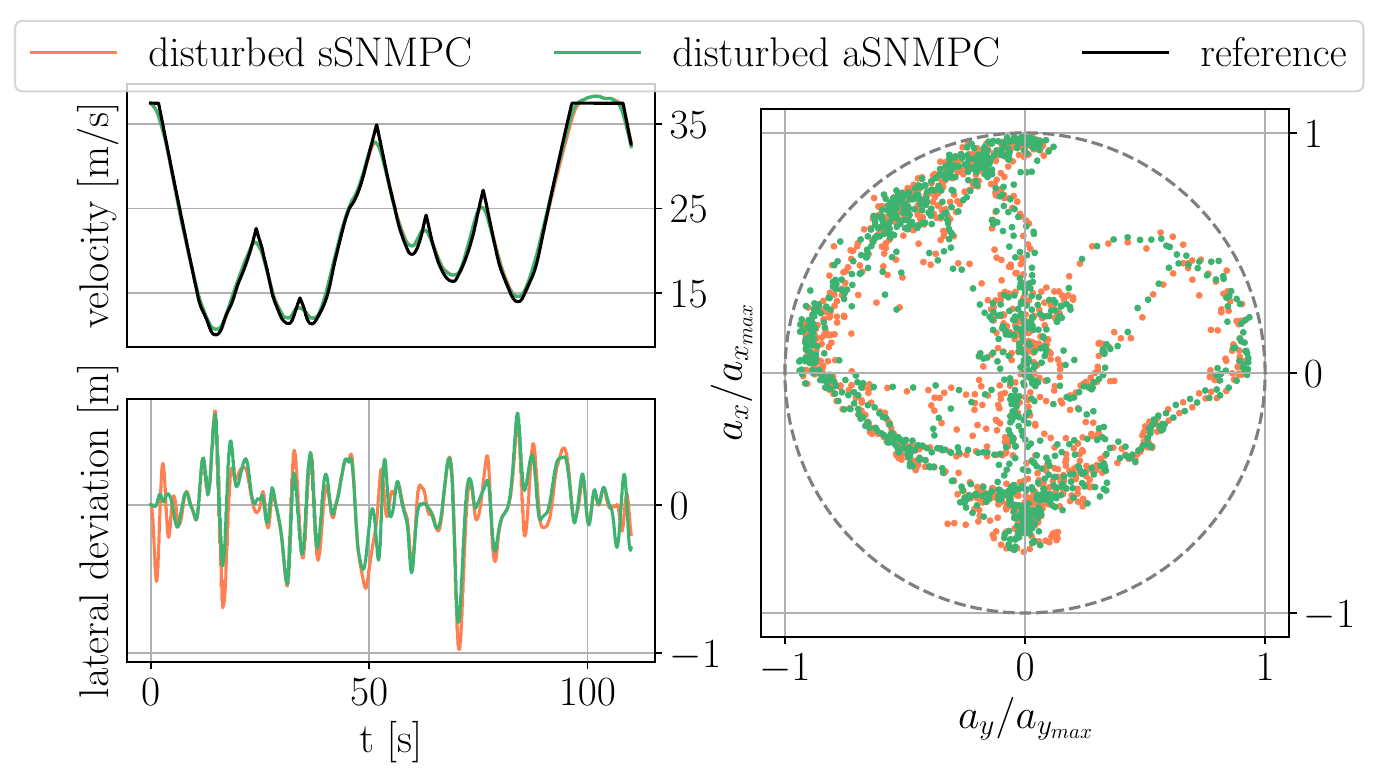}
\caption{Velocity deviation and lateral deviation and the gg-diagram plots showcasing the closed-loop performance on Monteblanco racetrack with reference velocity up to $37.5\unit{\meter\per\second}$.}
\label{fig:benchmark_disturbed_nmpcs_gg}
\end{figure}

\begin{figure}[H]
\includegraphics[width=.48\textwidth]{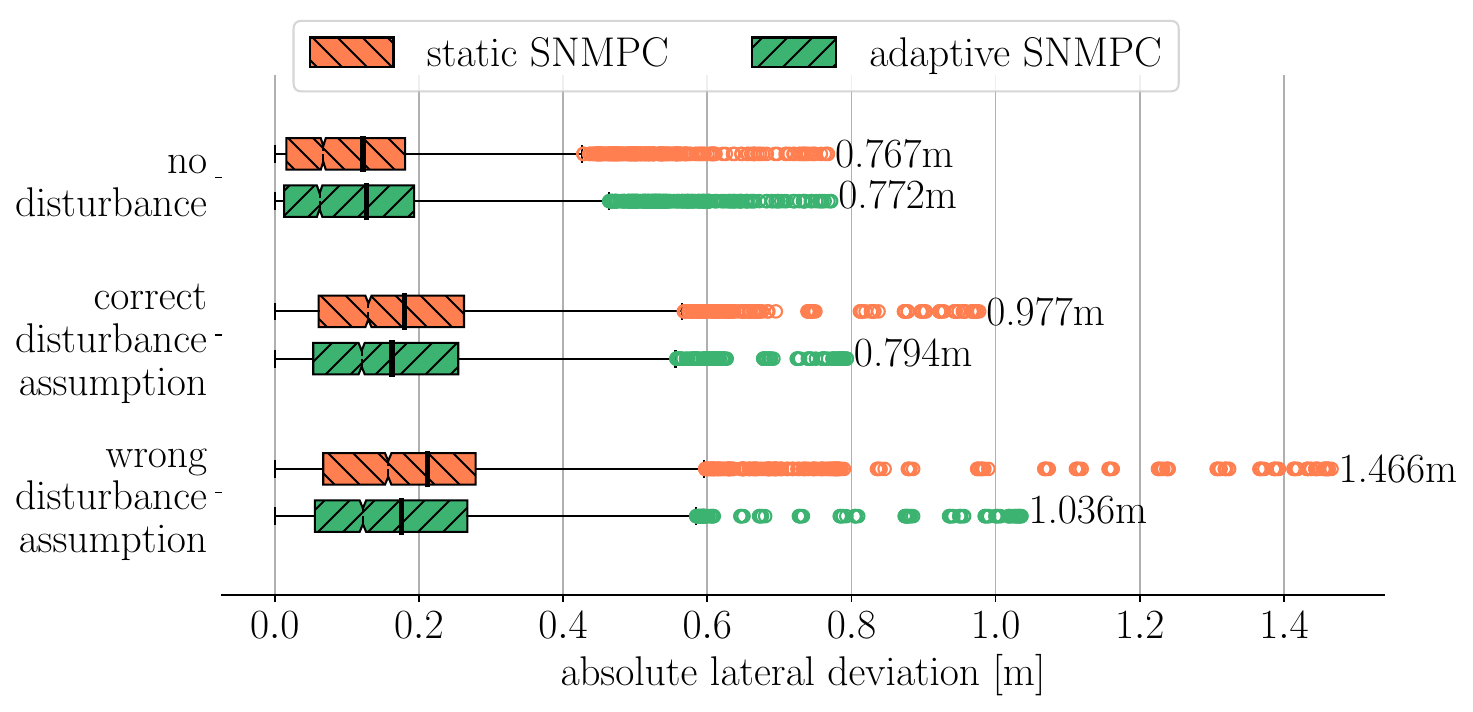}
\caption{Monteblanco racetrack: Effect of the disturbance on the sSNMPC and aSNMPC w.r.t. the absolute lateral deviation.}
\label{fig:benchmark4nmpcs}
\end{figure}

\begin{figure}[H]
\includegraphics[width=.48\textwidth]{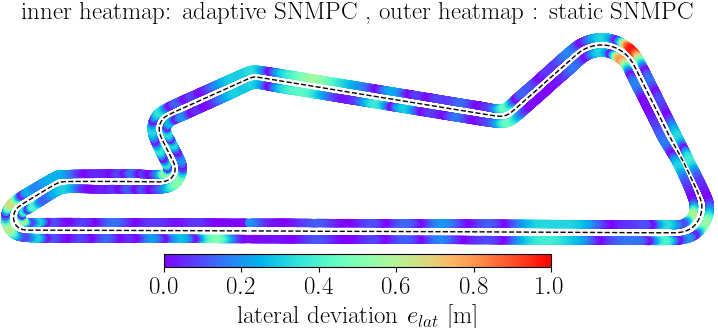}
\caption{Heatmap of lateral deviation for the aSNMPC and sSNMPC over one Monteblanco lap.}
\label{fig:lateraldistribution}
\end{figure}

Once the agent is trained, we run an evaluation simulation affected by time-varying external Gaussian disturbances $\boldsymbol{\sigma}_{w}^{\text{sim}}= [\sigma_x,\sigma_y,\sigma_{\psi},\sigma_{v,\text{lon}},\sigma_{v,\text{lat}},\sigma_{\dot{\psi}},\sigma_{\delta_f}]^T$, where the standard deviations are altered every $30\unit{\second}$, i.e. every  1500 SNMPC steps, and are randomly sampled from the ranges as in Tab.\ref{tab:std ranges}. \\
Figure \ref{fig:benchmark_disturbed_nmpcs_gg} provides a comparative assessment of sSNMPC and aSNMPC under the influence of strong disturbances. Both controllers exhibit similar performance in following the reference velocity profile and satisfying the nonlinear acceleration constraints (Eq.\ref{eq:combined constraints}). However, a notable difference arises w.r.t. the lateral deviation.
In Figure \ref{fig:benchmark4nmpcs}, we evaluate both MPCs in three different scenarios, focusing on absolute maximum lateral deviation. Minimizing this metric is crucial for ensuring lane-keeping and avoiding collisions with other road participants. 
While both SNMPCs perform similarly under nominal conditions, i.e. no disturbances, aSNMPC significantly outperforms sSNMPC when exposed to time-variant disturbances (see Tab.\ref{tab:std ranges}). With accurate disturbance assumptions, aSNMPC achieves a \textbf{18.73\%} improvement over sSNMPC in limiting maximum lateral deviation.\\
Importantly, aSNMPC demonstrates robust behavior by only showcasing a marginal \textbf{2.8\%} degradation in maximum deviation when significant disturbances are present, compared to its optimal performance in the absence of disturbances.
Conversely, when incorrect disturbance assumptions are made, i.e. $\boldsymbol{\sigma}_{w}^{\text{SNMPC}} \neq \boldsymbol{\sigma}_{w}^{\text{sim}}$, aSNMPC still outperforms sSNMPC by a significant \textbf{29.3\%}, underscoring its robustness against outliers and extreme events. However, aSNMPC's performance declines by \textbf{23.3\%} compared to correct disturbance assumptions, highlighting the critical importance of accurate disturbance assumptions and the need for a dedicated disturbance estimation unit when deploying SNMPC on a vehicle.
\\
Figure \ref{fig:lateraldistribution} illustrates the distribution of lateral deviations during the lap, highlighting that significant deviations mainly occur during curves. 
\subsection{Analyzing Reinforcement Learning agent decisions}
\label{subsec:RL decisions}
In Figure \ref{fig:kappa_Tu_heatmap}, we observe and analyze the decision-making patterns of the RL agent during a full lap. 
Before entering a curve while decelerating on a straight road, the agent increases the UPH to its maximum, i.e. chooses to propagate uncertainty across the entire prediction horizon. It then reduces the robustification factor $\kappa$ until it reaches its minimum value, ensuring the SNMPC is not overly conservative and that the vehicle follows the reference trajectory with minimal deviation. \\
During a curve, the agent gradually increases $\kappa$ to a mid-range value, thereby slightly tightening the constraints. Also, the DNN gradually decreases $T_u$ and stabilizes it at approximately 1.5\unit{\second} after the curve to prevent infeasibility caused by significant lateral disturbances encountered on straight paths. \\
After a curve while accelerating on a straight road, the agent reduces both $\kappa$ and $T_u$, i.e. stops tightening constraints, allowing the vehicle to operate at its longitudinal acceleration limits.\\
While maintaining a constant velocity on a straight road, the agent increases $\kappa$ to tighten constraints and mitigate the risk of aggressive control.
\subsection{Enhancing feasibility with the aSNMPC}
\label{testingfeasibility}
\begin{figure}[t]
\includegraphics[width=.48\textwidth]{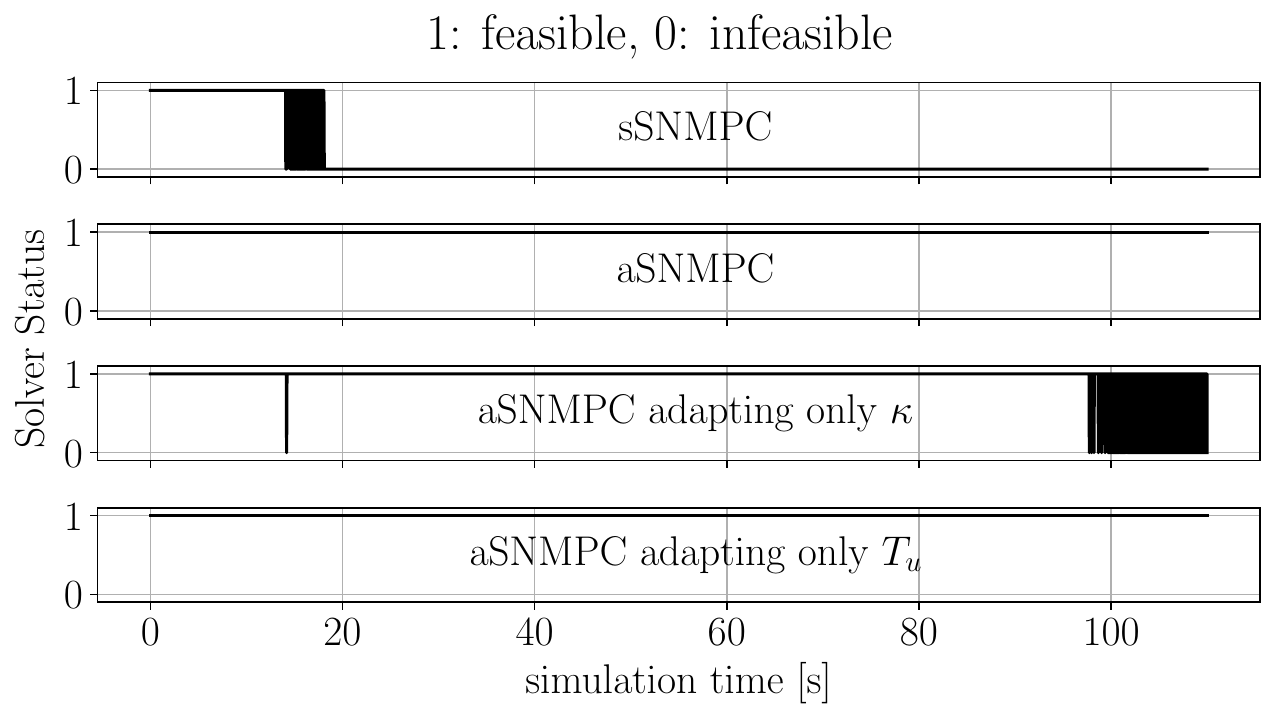}
\caption{Status of the solver for both SNMPC and aSNMPC under large time-varying external disturbances, black areas denote that the status oscillates with a high frequency.}
\label{fig:solverstatus_ASNMPC}
\end{figure}

We conduct simulations with larger disturbances than described in Tab.\ref{tab:std ranges}:
\begin{equation}
    \begin{aligned}
   &0.8\unit{\meter\per\second} \leq \sigma_{v\text{,lon}} \leq 1.5\unit{\meter\per\second}\\
    &0.7\unit{\meter\per\second} \leq \sigma_{v\text{,lat}} \leq 1.2\unit{\meter\per\second} \\ 
    &0.05\unit{\radian\per\second} \leq \sigma_{\dot{\psi}} \leq 0.08\unit{\radian\per\second}\\
    \end{aligned}
\label{eq:sim_dist_feasibility_setup}
\end{equation}
In Figure \ref{fig:solverstatus_ASNMPC}, we illustrate the solver's status during the simulation.
Notably, propagating uncertainties through the a fixed UPH, as in sSNMPC, leads to an infeasible problem when facing large external disturbances. However, adapting the UPH, as in RL driven aSNMPC, based on different driving tasks and uncertainties renders the problem feasible at each time step.
\begin{figure}[H]
\includegraphics[width=.48\textwidth]{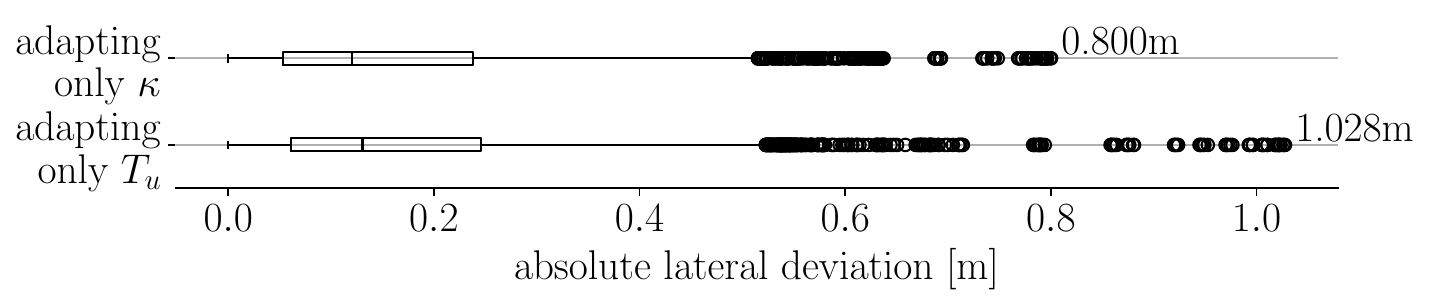}
\caption{Monteblanco racetrack: effect of the disturbance on two RL driven aSNMPC trained only to adapt $\kappa$ and $T_u$ respectively w.r.t. the absolute lateral deviation.}
\label{fig:boxplots_only_kappa_uph}
\end{figure}
This further showcases the importance of adapting UPH in a dynamic environment compared to SNMPC with fixed UPH.
\subsection{Impact of learning the robustification factor $\kappa$ and the Uncertainty Propagation Horizon $T_u$}
To demonstrate the influence of each parameter, we modify the action space (as discussed in Sec.\ref{subsec:action space}) and train two distinct RL agents. Each agent is designed to learn and adapt only a specific parameter:
\begin{enumerate}
    \item RL Agent 1: trains and learns only the constraints robustification factor $\kappa$, i.e., aSNMPC adapting only $\kappa$. 
    \item RL Agent 2: trains and learns only the UPH $T_u$, i.e. aSNMPC adapting only $T_u$. 
\end{enumerate}
Figure \ref{fig:boxplots_only_kappa_uph} illustrates the absolute lateral deviation evaluation for each agent. We notice that adapting the robustification factor $\kappa$ affects more the performance than adapting the UPH $T_u$ and ensures a $18.11\%$ improvement compared to the Static SNMPC (Fig.\ref{fig:benchmark4nmpcs}), as adapting $\kappa$ leads to less conservatism.\\
Conversely, as depicted in Fig.\ref{fig:boxplots_only_kappa_uph}, adapting only $\kappa$ results in infeasibility similar to sSNMPC. In contrast, adapting solely the UPH, $T_u$, makes the problem feasible, underscoring the importance of UPH in enhancing SNMPC's feasibility \cite{zarrouki2023stochastic}.
\subsection{Generalization and robustness}
\label{subsec: generalization}
\begin{figure}[h]
\includegraphics[width=.48\textwidth]{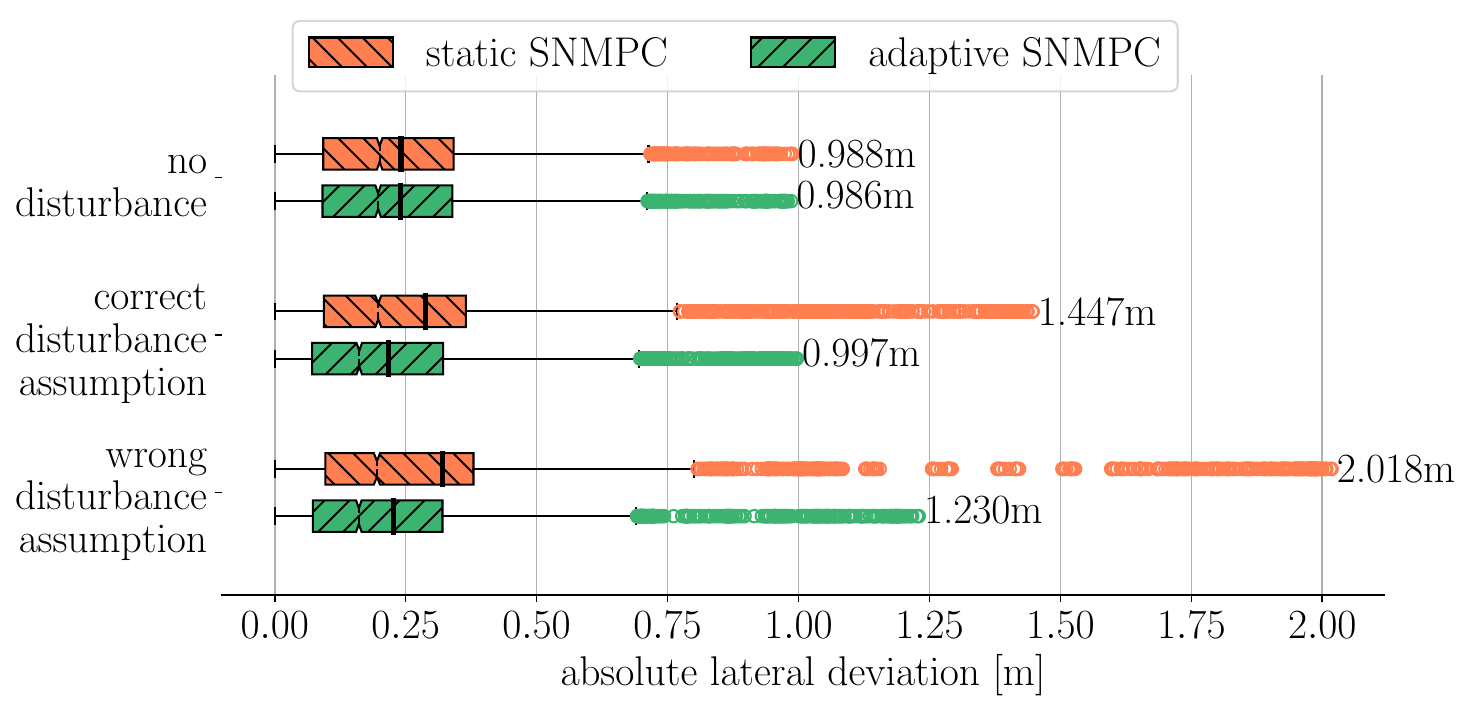}
\caption{Inexperienced Modena racetrack: 
effect of the disturbance on the aSNMPC and sSNMPC w.r.t. the absolute lateral deviation.
during \unit{110\second} = 5500 simulation steps subject to varying disturbance ranges.}
\label{fig:sim_result_modena}
\end{figure}

\begin{figure}[h]
\includegraphics[width=.48\textwidth]{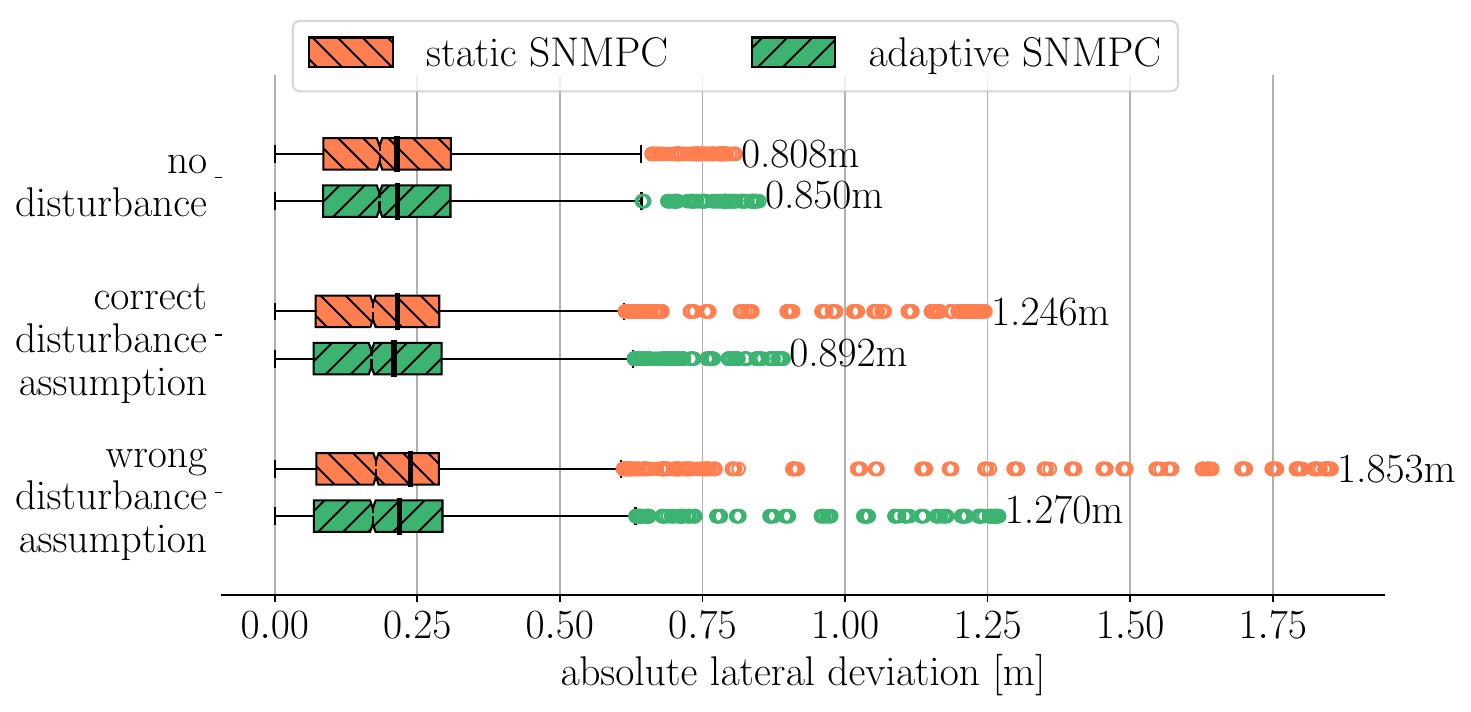}
\caption{Inexperienced Las Vegas Motor Speedway (LVMS): effect of the disturbance on the aSNMPC and sSNMPC w.r.t. the absolute lateral deviation.
during \unit{110\second} = 5500 simulation steps subject to varying disturbance ranges.} 
\label{fig:sim_result_lvms}
\end{figure}

The RL agent is only trained on Monteblanco racetrack. To assess its adaptability and generalization capabilities, we challenge the agent to adapt SNMPC parameters in two previously unexplored tracks while exposed to significant disturbances: Modena (Fig. \ref{fig:sim_result_modena}) and Las Vegas Motor Speedway (LVMS) (Fig.\ref{fig:sim_result_lvms}). We notice that the RL driven aSNMPC compared to the sSNMPC consistently achieves performance comparable to its training on the Monteblanco racetrack as described in Sec.\ref{subsec:traj_foll_performance}. On both unfamiliar tracks, our aSNMPC outperforms the sNMPC under the different disturbance settings. Notably, when we assume an accurate disturbance assumption, our aSNMPC showcases an improvement of $31\%$ on Modena and $28.4\%$ on LVMS compared with the standard SNMPC. When disturbance assumptions are inaccurate, the disparity widens, resulting in an impressive 39\% improvement on Modena and 31\% on LVMS.
\section{Conclusions and Future Work}
\label{sec:conclusion}
To enhance the feasibility of Stochastic Nonlinear Model Predictive Control (SNMPC) and to improve its closed-loop performance, we design a look-ahead Deep Reinforcement Learning (RL) agent to automatically learn and adapt two major SNMPC parameters:
the nonlinear constraints robustification factor $\kappa$ and Uncertainty Propagation Horizon (UPH) $T_u$. Leveraging the current kinematic state, disturbance assumptions, performance metrics, and future reference trajectory, the RL agent anticipates upcoming control tasks and dynamically selects the most suitable SNMPC configuration to optimize predefined objectives. \\
Our experimental findings reveal that the robustification factor $\kappa$ affects the closed-loop performance and its adaptation improves it by being less conservative than the Static SNMPC. On the other hand, adapting the UPH length, $T_u$, has a substantial effect on feasibility, rendering previously infeasible static SNMPC problems feasible when subjected to strong disturbances.\\
In the context of motion control for autonomous vehicles following a raceline trajectory at speeds of up to $37.5\unit{\meter\per\second}$, our Adaptive SNMPC (aSNMPC) exhibits a significant improvement in limiting maximum lateral deviation, with a 18.73\% enhancement under accurate disturbance assumptions and a substantial 29.3\% improvement when inaccurate disturbance assumptions are considered, showcasing its robustness against extreme events.\\
An analysis of the RL agent's decisions reveals that its actions are contextually driven by the current vehicle state and upcoming driving situations, such as acceleration, deceleration, or maintaining velocity on straight sections, both before, during, and after a curve. The trained Deep Neural Network (DNN) automatically selects the most suitable parameters, striking a balance between conservativeness regarding nonlinear constraint tightening, sensitivity to uncertainty propagation, feasibility, and closed-loop performance objectives.\\
Our experimental results validate the robustness of our Deep RL-driven aSNMPC, indicating that it avoids overfitting to the training conditions. In fact, our trained aSNMPC consistently outperforms the sSNMPC when facing unseen racetracks, demonstrating its ability to effectively handle disturbances in various environments. 

Nevertheless, as evident in Figures \ref{fig:benchmark4nmpcs}, \ref{fig:sim_result_modena}, and \ref{fig:sim_result_lvms}, it is apparent that both Static and Adaptive SNMPC performance experiences significant degradation when the disturbance assumptions are inaccurate. To this end, we propose two key directions for future research. First, we recommend extending the SNMPC framework with an online disturbance estimation unit. Second, we suggest incorporating the emulation of inaccurate disturbance assumptions during the agent's training phase to enhance its adaptability in challenging scenarios. 
\bibliographystyle{IEEEtran}
\bibliography{literatur} 

\end{document}